\PassOptionsToPackage{dvipsnames}{xcolor}
\documentclass[conference,9pt]{IEEEtran}
\IEEEoverridecommandlockouts
\usepackage{cite}
\usepackage{amsmath,amssymb,amsfonts}
\usepackage{algorithmic}
\usepackage{graphicx}
\usepackage{textcomp}
\usepackage{tikz}
\usepackage{tikzscale}
\usepackage{pgfplots}
\usepackage{optidef}
\usepackage{wrapfig}
\usepackage{pgfplots}
\usepackage[justification=centering]{caption}
\usepackage{etoolbox}
\usepackage{placeins}
\usepackage{pgfmath}
\usepackage[compact]{titlesec}
\titlespacing*{\section}{0pt}{0.1\baselineskip}{0.2\baselineskip}
\expandafter\def\expandafter\normalsize\expandafter{%
    \normalsize
    \setlength\abovedisplayskip{5pt}
    \setlength\belowdisplayskip{5pt}
    \setlength\abovedisplayshortskip{5pt}
    \setlength\belowdisplayshortskip{5pt}
}
\usetikzlibrary{positioning,fadings,through}
\pgfplotsset{compat=1.15}
\usepgfplotslibrary{groupplots}
\usepgfplotslibrary{colorbrewer}
\usepgfplotslibrary{fillbetween}
\usetikzlibrary{patterns}
\usetikzlibrary{circuits.ee.IEC}
\usetikzlibrary {shapes.geometric}

\def\BibTeX{{\rm B\kern-.05em{\sc i\kern-.025em b}\kern-.08em
    T\kern-.1667em\lower.7ex\hbox{E}\kern-.125emX}}
\pgfplotsset{axis background/.style={fill=gray!25}, axis line style={draw=none}, tick style={draw=none}, ymajorgrids, xmajorgrids, x grid style={white}, y grid style={white}}
\begin{document}

\title{LRMP: \underline{L}ayer \underline{R}eplication with \underline{M}ixed \underline{P}recision for Spatial In-memory DNN Accelerators
\thanks{This work was  supported in part by the U.S. Department of Energy, Office of Science, for support of microelectronics research, under contract number DE-AC0206CH11357, and in part by the National Science Foundation under grant CCF-2106964}
}

\author{\IEEEauthorblockN{Abinand Nallathambi\IEEEauthorrefmark{1},
Christin David Bose\IEEEauthorrefmark{1},
Wilfried Haensch\IEEEauthorrefmark{2},
Anand Raghunathan\IEEEauthorrefmark{1}}
\IEEEauthorblockA{\IEEEauthorrefmark{1}School of Electrical and Computer Engineering, Purdue University, West Lafayette, IN}
\IEEEauthorblockA{\IEEEauthorrefmark{2}Material Science Division, Argonne National Laboratory, Argonne, IL}
}
\maketitle

\begin{abstract}
In-memory computing (IMC) with non-volatile memories (NVMs) has emerged as a promising approach to address the rapidly growing computational demands of Deep Neural Networks (DNNs). Mapping DNN layers spatially onto NVM-based IMC accelerators achieves high degrees of parallelism. However, two challenges that arise in this approach are the highly non-uniform distribution of layer processing times and high area requirements. We propose LRMP, a method to jointly apply layer replication and mixed precision quantization to improve the performance of DNNs when mapped to area-constrained NVM-based IMC accelerators. LRMP uses a combination of reinforcement learning and integer linear programming to search the replication-quantization design space using a model that is closely informed by the target hardware architecture. Across five DNN benchmarks, LRMP achieves 2.8-9$\times$ latency and 11.8-19$\times$ throughput improvement at iso-accuracy.

\end{abstract}


\section{Introduction}
\label{sec:intro}

Deep Neural Networks (DNNs) have come to dominate the field of machine learning, and achieve state-of-the-art performance on a variety of complex tasks. However, the advancements in their capabilities have come at the cost of a steep growth in model sizes and computational complexity. Researchers have developed specialized digital accelerator architectures (\cite{tpu,cerebras,diannao}) and methodologies like quantization and pruning (\cite{liang2021pruning}) that attempt to strike better tradeoffs between cost and functional performance. These implementations are, however, fundamentally limited by the memory bottleneck, as memory accesses are significantly more expensive than digital compute. In-Memory Computing (IMC) is a computing paradigm where DNN computations are performed within memory arrays, potentially alleviating the memory bottleneck.

In neural networks, the multiplication of input vectors with weight matrices is an elementary operation. In-memory computing is a natural fit for vector-matrix multiplication (VMM) operations. IMC systems have been designed and prototyped with various memory technologies, including SRAM (\cite{kang2020deep, yin2020xnor,zhang2017memory}), DRAM (\cite{gao2019computedram}), and emerging NVMs such as RRAM (\cite{isaac,prime,pipelayer}), PCM (\cite{burr2015experimental,narayanan2021fully,khaddam2021hermes}) and STT-MRAM (\cite{yan2018celia,sttJain}). We focus on IMC with NVMs, where
the data is programmed as the resistance, or conductance, of the memory device. To perform a VMM operation using such a memory array, the weights can be programmed into the memory and the input vector can be presented by simultaneously activating the corresponding wordlines. The current that flows through each device is the product of the device conductance (weight) and the wordline voltage (input). These currents naturally sum up at each bit line. These behaviors are dictated by Ohm's and Kirchoff's laws. The currents at the bit lines thus produce the dot product of the input vector and the weight matrix. These bit line currents can then be digitized using analog to digital converters for further computation. 

Emerging non-volatile memories have a lot of desirable qualities that make them a good fit for in-memory computing. Their high density means that large amounts of data can be stored, and their non-volatility eliminates  the need for continuous power or refresh. However, the energy and latency costs of writing NVMs are high. They also have limited endurance, which means that they can only be reprogrammed a limited number of times. These factors make NVMs suitable for weight-stationary inference architectures where all the weights are programmed spatially across the chip and activations flow through and get processed by the appropriate arrays. A consequence of this approach is that the required area scales with the size of the network. While NVM arrays are compact, the peripherals required for IMC (ADCs, DACs and digital logic) imply that the required area can be quite large. For example, a 25mm$^2$ RRAM-based IMC chip (\cite{mchang}) in 40nm contains 288 256x256 RRAM tiles, whereas implementing even the modest ResNet18 DNN with 8-bit weights requires $\sim$ 1600 tiles.

The processing speed achieved by IMC implementations also depends greatly on how the DNNs are mapped onto them. It is known that convolution is naturally inefficient on IMC fabrics and the convolution layers can become the bottlenecks to latency and throughput. Layer replication (\cite{rasch2019rapa}) is a simple technique wherein bottleneck layers are replicated to facilitate data parallelism and thus, improve performance. However, layer replication is not a trivial optimization. Finding the resources to replicate the layers, choosing the right layers to replicate, and
the number of times to replicate the chosen layers, are all non-trivial decisions that involve complex tradeoffs. Also, with growing model sizes, improving performance of IMC architectures with layer replication becomes challenging since it exacerbates the required hardware resources.

In this work, we propose Layer Replication through Mixed Precision quantization (LRMP): an approach for optimizing the performance of DNNs through joint layer replication and quantization. Quantization frees up resources and the freed-up resources are re-used to selectively replicate layers to improve performance. To summarize, the contributions of this work are as follows:

\begin{itemize}
    \item We propose LRMP, a synergistic methodology of performing heterogeneous quantization and layer replication to improve the latency and throughput of area-constrained IMC-based neural network accelerators.
    \item We present a joint-optimization approach with a deep reinforcement learning based framework that selects the precision for each layer in the network with regard to performance and accuracy, and a linear programming based approach to selectively replicate layers by reutilizing the conserved resources in the IMC hardware to improve performance.
    \item We evaluate the LRMP framework on a benchmark suite of convolutional and fully-connected neural networks and achieve $2.8-9\times$ latency improvement and $11.8-19\times$ throughput improvement, at near iso-accuracy.
\end{itemize}

The rest of the paper is organized as follows. We describe the process of mapping a neural network layer in an IMC system and discuss the implications of precision on resource requirements and latency in Section \ref{sec:prelim}. We motivate the synergy between mixed precision and layer replication using an illustrated example in Section \ref{sec:motivation}. We present the details of the LRMP framework in Section \ref{sec:concepts}. We describe our experimental setup in Section \ref{sec:method} and present our results in Section \ref{sec:results}. We discuss the contributions of our work in the context of existing related works in Section \ref{sec:related}. Finally, we conclude the paper in Section \ref{sec:conclusion}.
\section{Preliminaries}
\label{sec:prelim}

Vector-matrix multiplication (VMM) is an elementary operation in the evaluation of neural networks. In this section, we describe how a weight matrix can be mapped to multiple crossbar tiles and how these crossbar tiles can collectively perform a multiplication operation between an input vector and a weight matrix to produce an output vector. We also describe the latency and the number of crossbar tiles required for such an implementation of VMM.

Convolutional layers represent a common layer configuration used in DNNs. They are composed of a three-dimensional weight tensor array sliding across a three-dimensional input tensor, producing an output value for each patch of overlap. Conventional layers are realized on IMC substrates by lowering the weight tensor to a two-dimensional matrix and the patches of the input tensor into a sequence of vectors. Then, the convolution output values can be produced by performing a sequence of VMMs. 

\begin{figure*}
    \centering
    \includegraphics{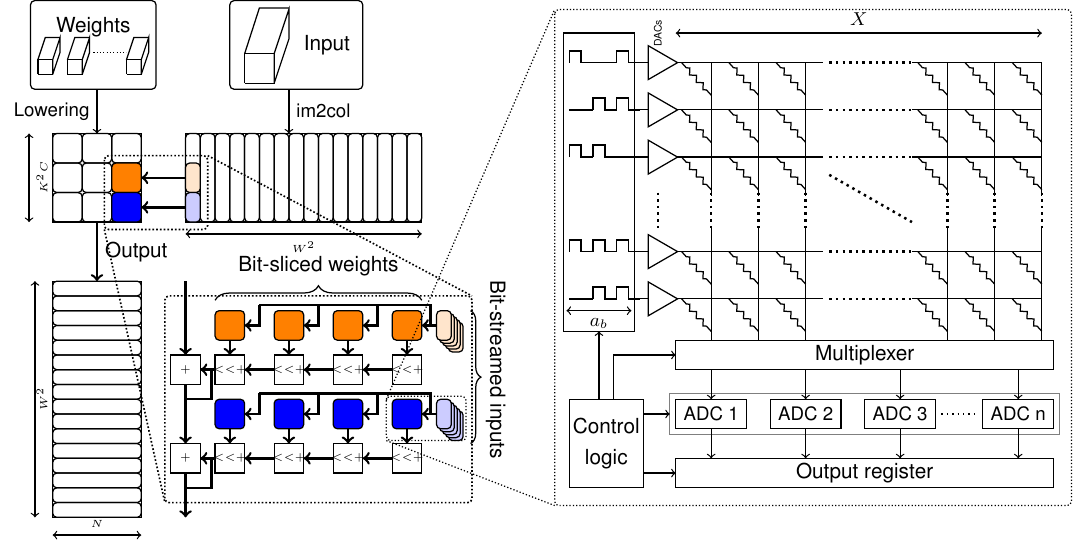}
    \caption{Functional description of a VMM operation using crossbar arrays}
    \label{fig:vmm}
\end{figure*}

Consider a convolution operation with $C$ input features, $N$ output features and a kernel size of $K$, producing an output features of dimension $W\times W$. The size of its lowered weight matrix is $K^2C\times N$. The input tensor is transformed into $W^2$ vectors of length $K^2C$. With a sequence of VMMs, $W^2$ output vectors of length $N$ are produced. It must be noted that the number of vectors can be quite high and it depends on the dimensions of the input, the filter kernel size $K$, the padding and the stride. For, example, in the first convolutional layer of the ResNet18 DNN, the input matrix has over 12,000 column vectors.

To map a convolutional layer to a crossbar, the weight tensor is first lowered to a two-dimensional matrix, as shown in Figure \ref{fig:vmm}. The weight matrix is then segmented into multiple sub-matrices of size $X\times X$, which denotes the size of the crossbar array or tile. To build a spatial architecture, each of these sub-matrices are mapped onto individual crossbar tiles. The number of tiles required to perform this spatial mapping is given by Equation \ref{eq:tile_count1}.

\begin{equation}
    \label{eq:tile_count1}
    \# \text{tiles}(K,C,N,X) = \left\lceil \frac{K^2C}{X} \right\rceil \times \left\lceil \frac{N}{X} \right\rceil
\end{equation}

As discussed in Section \ref{sec:intro}, crossbar arrays can be built with a variety of memory technologies. Also, these memory devices are designed to store a specific number of bits. The precision of the memory element is a matter of concern, as high precision elements have been shown to be more sensitive to process variations and conductance drift (\cite{wonbo}) and incur higher programming costs (\cite{perez}). Thus, low precision devices are desirable with regards to both accuracy and performance. 

The achievable precision of the memory device needs to be reconciled with the required logical precision of the weights. If the device precision ($s_b$) is less than the required weight precision ($w_b$), the weight sub-matrices can be sliced into groups of $s_b$ bits and then each slice can be mapped to a separate crossbar tile, as shown in Figure \ref{fig:vmm}. It must be noted that the digital outputs corresponding to the bit-slices of the weight matrix need to be appropriately shifted and added to produce the final output. The number of tiles required considering a bit-sliced mapping is given by Equation \ref{eq:tile_count2}.

\begin{equation}
    \label{eq:tile_count2}
    \# \text{tiles}(K,C,N,X,w_b,s_b) = \left\lceil \frac{K^2C}{X} \right\rceil \times \left\lceil \frac{N}{X} \right\rceil \times \left\lceil \frac{w_b}{s_b} \right\rceil
\end{equation}

As discussed in Section \ref{sec:intro}, to perform a vector-matrix multiplication using crossbars, we must convert the input vector values to analog currents using a digital-to-analog converter (DAC). The output of the crossbar array is then converted to digital using an analog-to-digital converter (ADC). These peripheral circuits, especially ADCs, occupy a significant proportion of latency and, area and power budgets. The design choices of the number of ADCs per crossbar array, and ADC and DAC precisions are important considerations in the design of crossbar-based architectures. The number of ADCs can be chosen to be lesser than the number of columns and the ADCs can be time-multiplexed between multiple columns. In order to reduce the precision requirements of ADCs and DACs, we can stream the input vectors bit-by-bit and reduce the corresponding outputs with shift-add operations. The latency of performing VMMs required by a convolution layer in such a bit-streamed manner is given by Equation \ref{eq:latency1}.

\begin{equation}
    \label{eq:latency1}
    \text{lat}(W,a_b,X,n_{ADC},t_{tile}) = W^2 \times t_{tile} \times \left\lceil \frac{X}{n_{ADC}} \right\rceil \times a_b
\end{equation}

where $n_{ADC}$ is the number of ADCs per crossbar array, $t_{tile}$ is the time elapsed between presenting an input to the tile and the ADCs producing their output, $a_b$ is the number of bits required to represent the input vector values and $W^2$ is the number of vectors.

Thus, both the hardware requirements and latency of a crossbar-based architecture depend on the precision of the weights and activations of the neural networks mapped onto them.

\section{Motivation}
\label{sec:motivation}
In this section, we illustrate the impact of mixed precision on tile allocation and latency/throughput of evaluation. We also demonstrate the benefits of selectively replicating layers in a DNN.

\begin{figure*}
    \centering
    \includegraphics{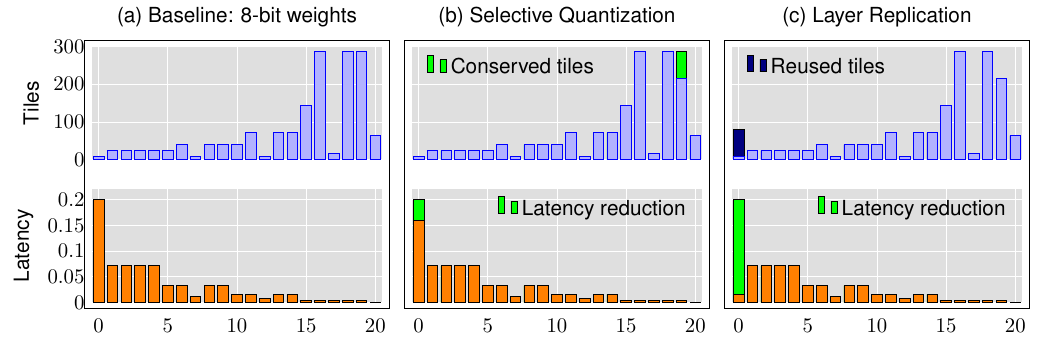}
    \caption{An experimental illustration of heterogeneous quantization and layer replication using ResNet18}
    \label{fig:motivation}
\end{figure*}

Let us consider the baseline implementation of ResNet18 with 8-bit weights and 8-bit activations. As defined by Equation \ref{eq:tile_count2}, the tile consumption of each layer in a spatial architecture depends on the size of the weight matrix and the precisions of weights ($w_b$) and memory device ($s_b$). As shown in Fig. \ref{fig:motivation}(a), we observe that different layers of the network have different latencies and tile costs, as defined by Equations \ref{eq:latency1} and \ref{eq:tile_count2}, respectively. In our evaluations, we use a device precision of 1-bit and a crossbar size of 256$\times$256.

By selectively reducing the precision of weights in certain layers, we can take advantage of the bit-sliced implementation and reduce the number of tiles required by that layer. These conserved tiles can be used to replicate bottleneck layers in a neural network to process multiple input vectors in parallel, resulting in a latency reduction. Similarly, by selectively reducing the precision of input vectors, we can reduce the number of bits to be streamed to the crossbar arrays, resulting in proportional reduction of latency.

Let us consider reducing the weight precision of a resource-intensive layer and the input precision of the bottleneck layer to 6-bits. As shown in Fig. \ref{fig:motivation}(b), we observe that 72 tiles are conserved. In addition, as explained by Equation \ref{eq:latency1}, the latency of the bottleneck layer is reduced resulting in an overall 5.7\% latency improvement and 1.33$\times$ throughput improvement.

If these newly freed-up tiles are used to naively replicate only the bottleneck layer, we can create 9 more copies of that layer. Thus, 10 input vectors of that layer can be processed in parallel, resulting in 25.5\% improvement in total latency and 2.34$\times$ improvement in throughput, as illustrated in Fig. \ref{fig:motivation}(c).

The above example illustrates that a trade-off exists between precision and latency in spatial IMC architectures. A few considerations that arise about this trade-off are:

\subsubsection*{How to choose the precision of each layer?}
When choosing the precision of each layer, we need to consider its impact on the tile consumption, overall latency, and accuracy. The weight precision affects the bit-slicing factor, which is only one of the factors that determines the tile consumption of a layer (Equation \ref{eq:tile_count2}). The other factors depend on the size of the weight matrix. Thus, it is important to choose the weight precision of each layer in a way that the number of tiles conserved is maximized. Similarly, the activation precision only affects the bit-streaming factor of Equation \ref{eq:latency1}. The other factor is the number of input vectors to be processed. Thus, it is important to choose the activation precision of each layer in a way that the latency is minimized. Moreover, reducing the activation/weight precision of any layer in a neural network has implications for the overall accuracy of the network. Thus, it is important to choose the precision of each layer in a way that the overall accuracy is not compromised.

\subsubsection*{Where to repurpose the conserved tiles?}
When choosing the replication factor of a layer, we need to consider its impact on the overall latency and tile consumption. The latency of a layer is a function of the number of input vectors and their precision, both of which can vary across the layers of a neural network. At the same time, the number of tiles required to replicate layers also varies based on the size of their weight matrices. Thus, it is important to choose the replication factor of each layer in a way that the utility of the conserved tiles is maximized, and the overall latency is minimized.

\section{LRMP Methodology}
\label{sec:concepts}

In this paper, we propose LRMP (Layer Replication through Mixed Precision), a framework that combines a reinforcement learning (RL) and linear programming (LP) to jointly optimize latency/throughput and accuracy of DNNs realized on IMC hardware fabrics. 
\begin{figure*}
    \centering
    \includegraphics{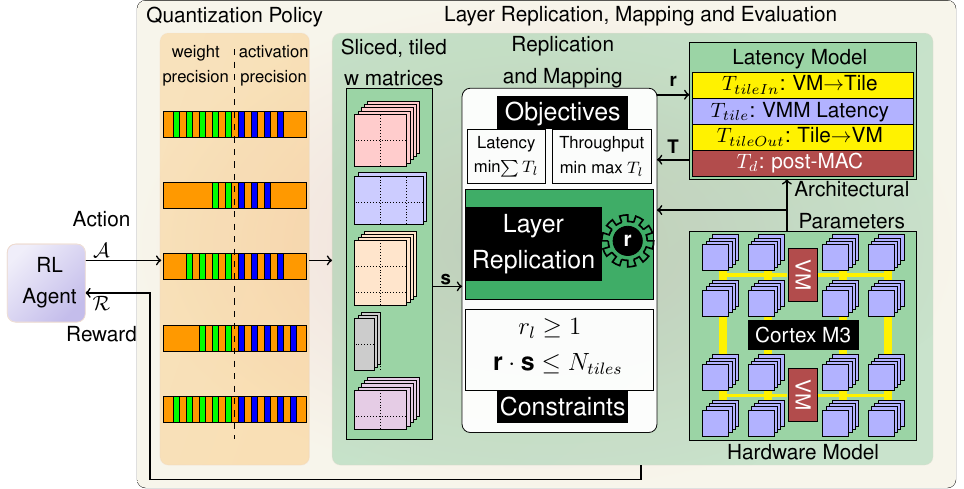}
    \caption{Overview of the proposed LRMP methodology.}
    \label{fig:blockdiagram}
\end{figure*}

As shown in Figure~\ref{fig:blockdiagram} LRMP is an iterative process with each iteration or episode consisting of two-steps: (1) an RL-agent choosing the precision of each layer in the DNN, and (2) an LP-based optimizer choosing the replication factors of each layer. After each episode, the latency/throughput achieved by the LP-based optimizer and the accuracy of the network are used to train the RL-agent. In the remainder of this section, we describe the hardware modeling of an RRAM-based IMC accelerator, and then discuss how linear programming and reinforcement learning can be employed in tandem to quantize and replicate layers to jointly optimize accuracy and performance metrics under a chip capacity constraint.

\subsection{Hardware Model}
LRMP is a joint optimization process that is designed to improve accuracy and latency/throughput achieved by DNNs on spatial in-memory accelerators. To estimate the performance metrics of evaluating DNNs on these spatial accelerators, we develop a simple and effective cost model that can be used to estimate latency and throughput of a DNN on the fly. The cost model is based on the compute-in-memory system developed by \cite{mchang}, which consists of a Cortex M3 microprocessor, two vector modules for digital compute, and 288 crossbar tiles of dimension 256$\times$256. Data transport is implemented using 8 lanes of 8-bit wide buses from the vector modules to the crossbar tiles and 8 lanes of 32-bit wide buses from the crossbar tiles to the vector modules. Each vector module has 8 lanes of parallel compute and 128KB of SRAM. The system is equipped with fine-grained power gating and each tile can be individually turned off. On account of the larger computer vision models used to benchmark the proposed approaches in this work, the cost model assumes a scaled-up version of this system with 5682 tiles and 40 vector modules, each with 64 lanes of parallel compute. Further details on the microarchitecture are provided in Section \ref{sec:method}.

The latency of evaluating a DNN layer ($T_l$) on this compute-in-memory system, described in Eqn. \ref{eq:layerLatency}, comprises of 4 factors: $T_{tileIn.l}$, which refers to the latency of transferring input vectors of layer $l$ from vector modules to the respective tiles. This latency is shared by eight 8-bit lanes, which are shared among 144 tiles. Similarly, $T_{tileOut.l}$ represents the latency of transferring output vectors of layer $l$ from crossbar tiles to the respective vector modules using eight 32-bit lanes shared by the same 144 tiles. The latency of performing Vector-Matrix-Multiplication (VMM) using crossbar tiles with temporally bit-streamed inputs and spatially bit-sliced weights of layer $l$ is represented by $T_{tile.l}$. Lastly, the post-VMM digital compute of layer $l$ performed by vector modules has a latency of $T_{d.l}$, which uses 64 lanes processing the output vectors of 144 tiles. It must be noted that each of these components is a function of the number of bits used to represent the input activations and weights of layer $l$.

\begin{equation}
        T_l = T_{tileIn.l} + T_{tileOut.l} + T_{tile.l} + T_{d.l} \label{eq:layerLatency} \\
\end{equation}

With the latency of evaluating a layer defined, the latency of evaluating a DNN is the sum of the latencies of its constituent layers. The latency of evaluating a DNN with $L$ layers is given by Eqn. \ref{eq:totalLatency}.
\begin{equation}
        T = \sum_{l=1}^{L} T_l \label{eq:totalLatency}
\end{equation}

The system is designed to operate with coarse-grained pipeline parallelism. Thus, the throughput of the system is defined by the maximum latency of any layer. The throughput of the system is given by Eqn. \ref{eq:throughput}.

\begin{equation}
        P = \frac{1}{\underset{l}{\max} \text{ } T_l} \label{eq:throughput}
\end{equation}

The model described above is used in LRMP to perform analysis and exploration of the quantization and replication design space.

\subsection{Optimizing layer replication using linear programming}

As discussed in Section \ref{sec:motivation}, tiles can be freed up by selectively quantizing layers based on their tile footprint and selectively replicating layers based on their latencies. When a layer is replicated, the number of tiles and vector modules allocated to that layer is increased. Thus, for the said layer: (1) the total bandwidth available for data transfer is increased; (2) the amount of digital compute allocated is increased; and (3) the number of tiles available for performing the required VMM operations is increased. This results in a linear reduction in the latency of evaluating the layer.

If there are $r_l$ instances of layer $l$, then the latency of evaluating the DNN is given by Eqn. \ref{eq:layerLatencyReplicated}.
\begin{equation}
    T = \sum_{l} \frac{1}{r_l} (T_{tileIn.l} + T_{tileOut.l} + T_{tile.l} + T_{d.l}) \label{eq:layerLatencyReplicated}
\end{equation}

Given a mixed precision quantization scheme, the total number of tiles required to have one instance of each layer ($s_l$) is given by Eqn. \ref{eq:tile_count2}. The total latency $T$ can be optimized by carefully choosing the layer replication factors $\textbf{r}$. The process of choosing the replication factors is naturally constrained by the total number of tiles available in the system ($N_{tiles}$). This can be formulated as a constrained optimization problem.

\begin{mini*}|l|[3]
    {\textbf{r}}{\sum_{l} \frac{1}{r_l} (T_{tileIn.l} + T_{tileOut.l} + T_{tile.l} + T_{d.l})}{}{}
    \addConstraint{r_l \geq 1}
    \addConstraint{\sum_l (r_l*s_l) \leq N_{tiles}}
\end{mini*}

The constraints ensure that there is at least one instance of each layer and the total number of allocated tiles doesn't exceed the number of tiles available ($N_{tiles}$). Since $s_l$ is constant for a given quantization scheme, the constraints are linear. However, the objective function is non-linear. 

As defined by Eqn. \ref{eq:throughput}, the throughput of the system is the inverse of the maximum latency across all layers. Thus, to maximize throughput, we need to minimize the maximum latency across all layers. Optimizing for throughput is, thus, a min-max problem. The optimization problem can therefore be re-written as follows.

\begin{mini*}|l|[3]
    {\textbf{r}}{M}{}{}
    \addConstraint{\frac{1}{r_l} (T_{tileIn.l} + T_{tileOut.l} + T_{tile.l} + T_{d.l}) \leq M}
    \addConstraint{r_l \geq 1}
    \addConstraint{\sum_l (r_l*s_l) \leq N_{tiles}}
\end{mini*}

We introduce a dummy variable $M$ and reformulate the optimization problem to minimize $M$, while ensuring that the latency of each layer does not exceed $M$. This ensures that the maximum latency across all layers is minimized. While the objective function of this formulation is linear, the first constraint is not. 

These optimization problems are not automatically linear. However, we can employ linearization techniques (\cite{linearize}) to reformulate the optimization problem with linear constraints and objective function. Then, we solve the reformulated problem using an LP solver.

\subsection{Constraining the action space with performance budgets}
The reinforcement learning framework used in this work is based on the work by \cite{haq}, which imposes a performance cost constraint on the action space of the RL agent. If the quantization policy prescribed by the RL agent does not meet the performance targets, it is modified by decreasing the bitwidths until the performance targets are met. While this approach is effective, it does not provide any insight into the tradeoffs between accuracy and performance. We restructure this approach to explore the tradeoffs between accuracy and performance by exponentially tightening the performance budget. This results in the RL agent exploring the space of quantization policies to not just meet a performance budget but also achieve better performance metrics.

\subsection{Rewarding the RL agent with accuracy and performance metrics}
In each episode of exploration, the RL agent is rewarded based on the quality of the quantization policy it prescribes. \cite{haq} rewarded the RL agent based on the accuracy of the quantized DNN. In this work, we optimize the performance of the quantized DNN by using the layer replication technique. Thus, to achieve joint optimization, the RL agent is rewarded based on the accuracy and performance of the quantized DNN. The reward function is given by Eqn. \ref{eq:reward}.

\begin{equation}
    \mathcal{R} = \lambda \times (acc_{quant} - acc_{original}) + \alpha \times (1 - T_{quant} / T_{original})
    \label{eq:reward}
\end{equation}

where, $acc_{quant}$ is the accuracy of the quantized DNN, $acc_{original}$ is the accuracy of the original DNN, $T_{quant}$ is the latency of the quantized DNN and $T_{original}$ is the latency of the original DNN when optimizing for latency. When optimizing for throughput, $T_{quant}$ and $T_{original}$ are latencies of the bottleneck layers of the respective DNNs. The hyperparameters $\lambda$ and $\alpha$ control the relative importance of accuracy and performance in the reward function. The reward function is designed to encourage the RL agent to prescribe quantization policies that result in a quantized DNN that is optimized to balance accuracy and speed. 
\section{Experimental Methodology}
\label{sec:method}

\subsection{Microarchitectural details}

As described in Section \ref{sec:concepts}, this work is based on a scaled-up model of the compute-in-memory system fabricated by \cite{mchang}. The microarchitectural parameters are listed in Table \ref{tab:arch_params}. 

The system is built using 1T-1R RRAM eNVM technology, with a tile size of 256x256 and a total of 5682 tiles. The system also includes 40 vector modules, each of which contains 64 lanes of parallel digital compute and 128 KB of SRAM. Each tile is equipped with eight 4-bit Flash ADCs and 256 1-bit DACs. To prevent partial sum quantization and mitigate other non-idealities, only 9 rows are activated at a time. The system is clocked at 192 MHz. 

\begin{center}
{\small
\begin{tabular}{|c|c|}
    \hline
    \textbf{Parameter} & \textbf{Value} \\
    \hline
    eNVM & 1T-1R RRAM \\
    \hline
    Tile size & 256$\times$256 \\
    \hline
    No. of tiles & 5682 \\
    \hline
    No. of vector modules & 40 \\
    \hline
    Device precision & 1 bit \\
    \hline
    Row parallelism & 9 \\
    \hline
    DAC precision & 1 bit \\
    \hline
    Column parallelism & 8 \\
    \hline
    ADC precision & 4 bits \\
    \hline
    Avg. power per tile & 70 $\mu$W \\
    \hline
    Clock frequency & 192 MHz \\
    \hline
\end{tabular}
\captionof{table}{Microarchitectural parameters}
\label{tab:arch_params}
}
\end{center}

\subsection{Methods}

\subsubsection*{Reinforcement learning}
As described in Section \ref{sec:concepts}, the reinforcement learning framework used in this work is based on the hardware-aware quantization tool proposed by \cite{haq}. The method consists of two phases: exploration and finetuning. In the exploration phase, the agent explores the action space to find a good policy based on the performance budget and rewards provided, as described in Section \ref{sec:concepts}. The trajectory of the exploration phase is discussed in Section \ref{sec:rlTrajectory}.

After the exploration phase, the DNN is quantized with the mixed precision scheme found by the agent. In the finetuning phase, the DNN is trained with the quantized weights and activations to recover any accuracy lost to quantization.

\subsubsection*{Linear programming}
Given a quantization policy prescribed by the RL agent, the linear programming step is used to find the replication factors that optimize the performance of the system. Optimization objectives of both latency (latencyOptim) and throughput (throughputOptim) are implemented. The baseline for each network in the benchmark suite is the implementation with 8-bit weights and activations. Thus, the layer replication is done with a constraint that the total number of tiles used may be no more than the baseline. This is a design choice to ensure that performance is optimized while utilizing the same number of tiles as the baseline. 
An ablation study has been performed and described in Section \ref{sec:ablation} to show the effectiveness of our LRMP method with or without this design constraint. 

\subsection{Benchmarks}
The proposed LRMP approach has been evaluated on a set of DNN benchmarks trained on the ImageNet and MNIST datasets. The baseline of comparison for each benchmark is the implementation with 8-bit weights and activations. The benchmarks are listed in Table \ref{tab:benchmarks}, along with the number of tiles required by the baseline implementation. The MLP network is trained on the MNIST dataset, with 4 hidden layers of 1024, 4096, 4096 and 1024 neurons respectively. The residual networks are finetuned on the ImageNet dataset with pre-trained weights. It must be noted that, besides quantization, analog non-idealities such as noise, conductance drift, device-to-device variation etc. have not been modeled in this work. However, modeling these non-idealities (\cite{roy2021txsim,jain2020rxnn,lu2021neurosim}) and developing compensation techniques (\cite{meng2021temperature,rasch2019rapa,charan}) are areas of active and ongoing research and we believe these effects are not an impediment to the principal contributions of this work.

\begin{center}
{\small
\begin{tabular}{|c|c|c|}
    \hline
    \textbf{Benchmark} & \textbf{Dataset} & \textbf{$N_{tiles}$} \\ \hline
    MLP & MNIST & 3232 \\ \hline
    ResNet18 & ImageNet & 1602 \\ \hline
    ResNet34 & ImageNet & 2965 \\ \hline
    ResNet50 & ImageNet & 3370 \\ \hline
    ResNet101 & ImageNet & 5682 \\ \hline
\end{tabular}    
}
\end{center}
\captionof{table}{DNN benchmarks}
\label{tab:benchmarks}
\section{Results}
\label{sec:results}
In the sub-sections of this section, we first present the latency, throughput and energy improvements achieved by LRMP. We then present results that provide insights into the RL-based exploration process. We also show a layer-wise breakdown of how latencies are optimized and an ablation study that analyses the sensitivity of the layer replication methodology to area constraints.

\subsection{Latency and throughput improvements}
\begin{figure}[b]
    \centering
    \includegraphics[width=\columnwidth]{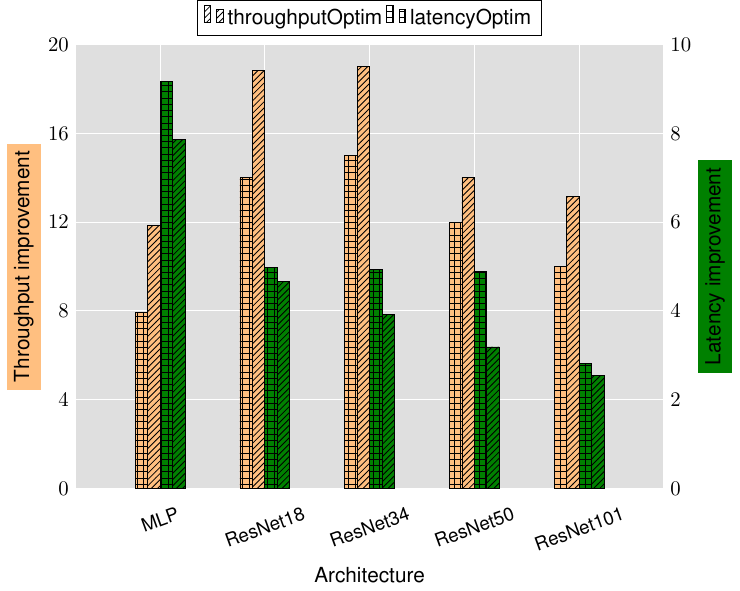}
    \caption{Latency and throughput improvements achieved by LRMP}
    \label{fig:r1}
\end{figure}

Fig. \ref{fig:r1} reports the latency and throughput improvements achieved by the LRMP framework. As explained in Section \ref{sec:method}, the improvements are reported with respect to fixed-precision baseline networks with 8-bit weights and activations. 

We observe 2.8-9$\times$ reduction in latency and 8-15$\times$ improvement in throughput while optimizing for latency ( denoted as latencyOptim) across the suite of benchmark DNNs. Similarly, we observe 11.8-19$\times$ improvement in throughput and 2.5-8$\times$ reduction in latency while optimizing for throughput (denoted as throughputOptim). These improvements are obtained with accuracy loss of less than 1\% after finetuning with the quantization policies determined by the exploration phase of LRMP.
\subsection{Energy improvements}
\begin{figure}
    \centering
    \includegraphics[width=\columnwidth]{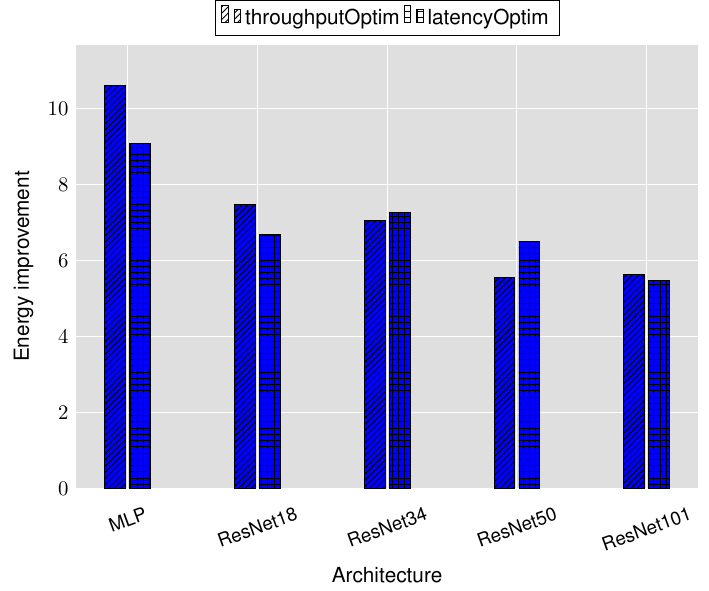}
    \caption{Energy improvements achieved by LRMP}
    \label{fig:r1b}
\end{figure}

Although LRMP explicitly optimizes for throughput or latency, it achieves energy improvements as a result of more efficient DNN execution on the IMC substrate. Fig. \ref{fig:r1b} shows the energy improvements achieved by LRMP. Energy consumption was modeled with three components: RRAM tile energy, memory accesses by the vector module and the leakage of the SRAMs. We observe 5.5-10.6$\times$ improvement in energy consumption while optimizing for throughput and 5.5-9$\times$ energy improvement while optimizing for latency.

\subsection{Studying joint optimization of accuracy and performance}
\label{sec:rlTrajectory}
\begin{figure}
    \centering
    \includegraphics[width=\linewidth]{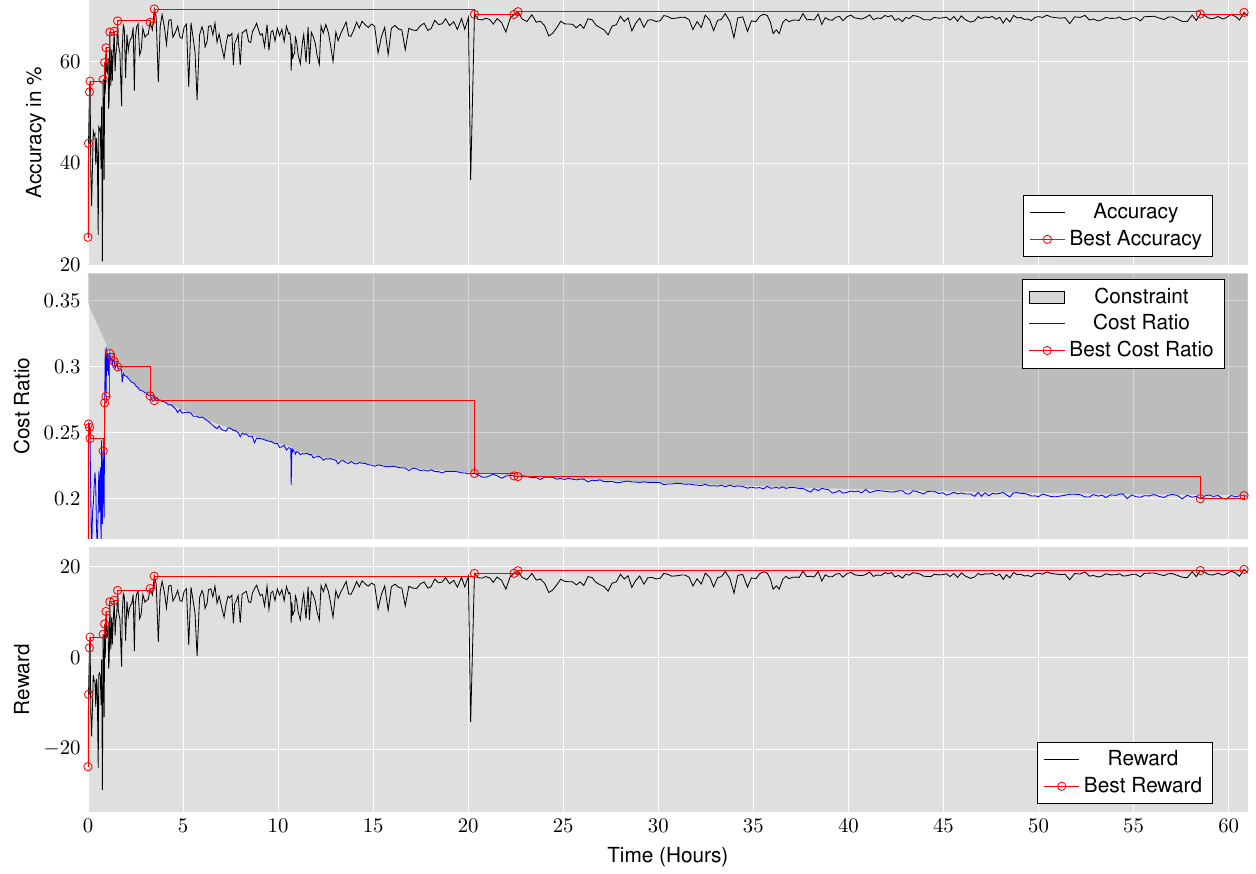}
    \caption{Trajectory of RL agent jointly optimizing ResNet18 for accuracy and latency}
    \vspace{-0.5cm}
    \label{fig:r33}
\end{figure}

As discussed in Section \ref{sec:concepts}, the proposed approach jointly optimizes for accuracy and performance by rewarding the RL agent with an affine combination of accuracy and performance metrics, and by continuously tightening the constraints placed on the action space. Fig. \ref{fig:r33} shows the trajectory of the RL agent performing latency optimization for ResNet18. The exploration is started with a lenient performance budget of 0.35$\times$baseline latency and exponentially tightened to 0.2$\times$baseline latency. Over the course of the exploration, the agent finds quantization policies that achieve upto 5$\times$ improvement in latency with layer replication while also improving the accuracy.

\subsection{Layer-wise latency breakdown}
\label{sec:r3}
\begin{figure*}
    \centering
    \includegraphics[width=\textwidth]{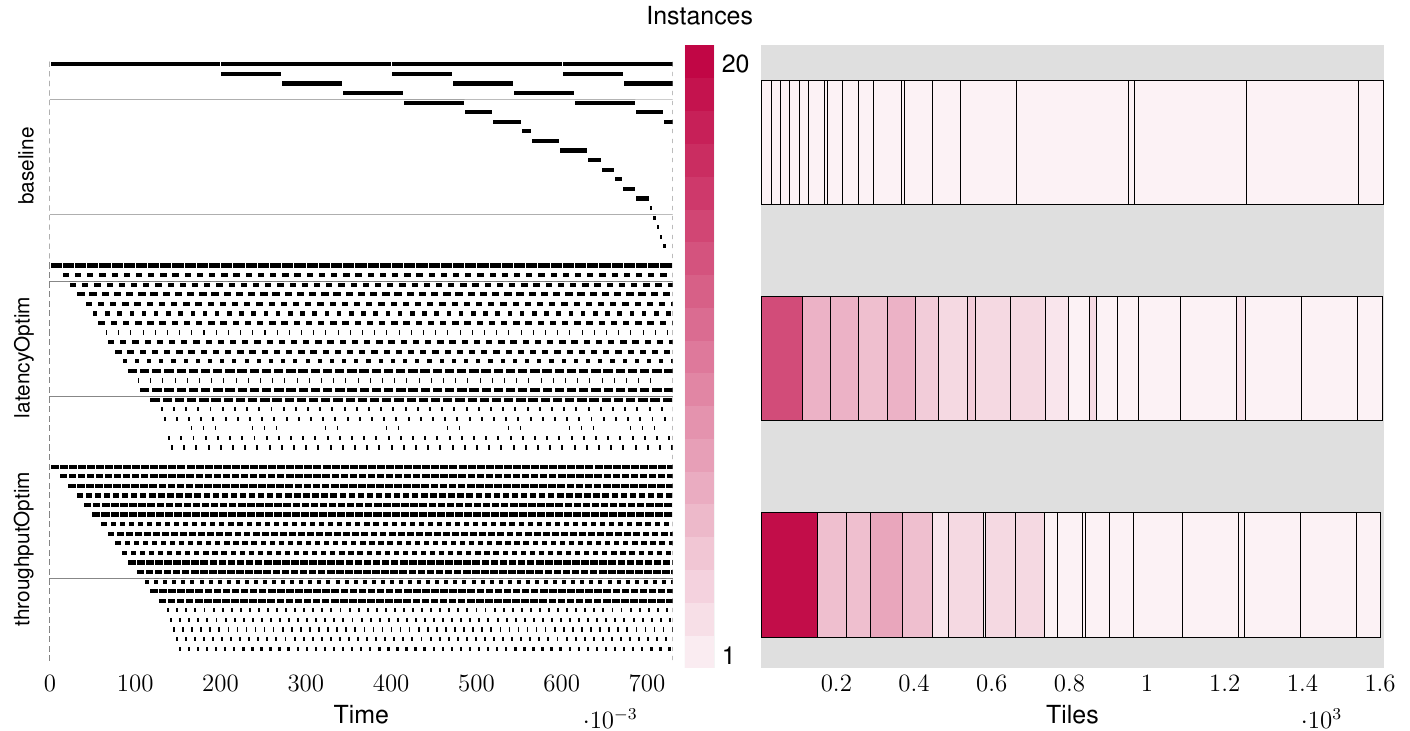}
    \caption{Layer-wise breakdown of latencies and tiles for ResNet18 for the baseline and while optimizing for latency (latencyOptim) and throughput (throughputOptim).}
    \vspace{-0.5cm}
    \label{fig:r2}
\end{figure*}

As discussed in Section \ref{sec:concepts}, the layer replication can be performed by optimizing for either latency or throughput. The two objectives have different implications on the latencies and tile consumptions of each layer and thus, latency and throughput outcomes for the overall network. Fig. \ref{fig:r2} shows the layer-wise breakdown of latencies and tiles for ResNet18 for the baseline implementation as well as the LRMP implementation while optimizing for latency and throughput.

In the baseline case, we observe that the latency of the network is bottlenecked by the first layer, which happens to consume very few tiles. When the layers are replicated for latency optimization (latencyOptim), the total latency is reduced by a factor of 5$\times$, while the latency of the bottleneck layer is reduced by 14$\times$ as 13 more copies of that layer are created. In the throughput optimization mode (throughputOptim), the total latency is reduced by a slightly smaller factor of 4.7$\times$, while the latency of the bottleneck layer is reduced by a larger factor of 19$\times$ as 18 more copies of that layer are created. This is understandable, because the bottleneck layer is solely responsible for determining throughput, while all layers contribute to latency.

\subsection{Analysis of sensitivity to chip area}
\label{sec:ablation}
As discussed in Section \ref{sec:method}, the layer replication methodology is performed with an area constraint based on the fixed precision baseline i.e., $N_{tiles}$ in the optimization constraints is equal to the number of tiles required by the fixed-precision 8-bit baseline network $baseline\_tiles$. 
We note that 
a different design choice, based on the chip area and power budgets, could result in the relaxation or tightening of this tiles constraint. 

\begin{figure}
    \centering
    \includegraphics[width=\columnwidth]{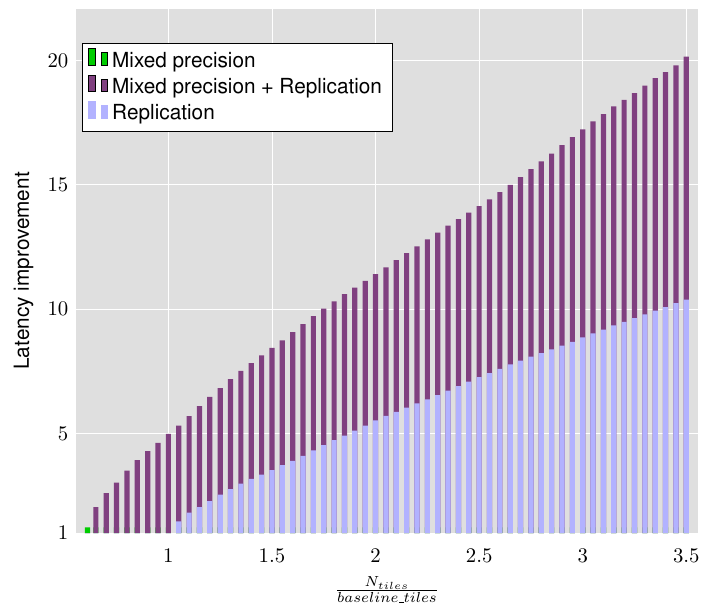}
    \caption{Latency and throughput improvements achieved by the proposed approach on ResNet18 with different area constraints.}
    \vspace{-0.5cm}
    \label{fig:r3}
\end{figure}

Fig. \ref{fig:r3} shows the sensitivity of the latency improvements achieved by LRMP to different area constraints for the ResNet18 DNN. 
We perform this analysis by setting $N_{tiles}$ to different ratios of $baseline\_tiles$ and using LRMP to perform only quantization, only replication, and joint quantization and replication. In other words, we study the behavior of LRMP by tightening the tiles constraint below the number of tiles required by the baseline or by relaxing the tiles constraint by making more tiles available in the system, while also using only one of the two optimization dimensions of LRMP.

Because of the model compression naturally achieved by mixed precision, with \emph{only} mixed precision, we achieve 18.5\% reduction in latency while using 39\% fewer tiles than the baseline. When we employ mixed precision \emph{and} layer replication, we observe latency reductions of 49\% while using 35\% fewer tiles than the baseline. 

We note that layer replication can be performed even without mixed precision, if more tiles are available. We employ \emph{only} layer replication with the baseline ResNet18 and observe 32\% reduction in latency while using 5\% more tiles than the baseline. It should be noted that when the tiles constraint is tightened, latency reductions are not possible without mixed precision, as there are not enough tiles for even a single copy of all the layers. Also, when all the tiles in the system are used, using mixed precision \emph{and} layer replication gives twice the latency improvement as compared to using only replication.


\section{Related Works}
\label{sec:related}

In this section, we discuss related previous work in the areas of quantization and pruning for DNN implementation on IMC hardware, as well as optimized mapping of DNNs to IMC substrates.
\subsection*{Quantization}
Quantization is the process of reducing the number of bits used to represent numbers, which naturally adds distortions in the form of quantization noise. Quantizing weights and activations of neural networks is a common technique to reduce the model size and improve performance of their implementations. Pruning is a technique that removes connections and neurons in a neural network to improve sparsity. These complementary techniques have been widely explored to optimize neural network implementations. Quantization and pruning techniques have also been applied specifically to the context of in-memory computing. \cite{peng} proposed a neural architecture search-based approach to perform mixed precision quantization in a crossbar-aware manner. \cite{kang} proposed a methodology to perform energy-aware quantization using a genetic algorithm. \cite{huang} proposed a methodology that performs quantization at the tile granularity powered by reinforcement learning. \cite{meng} proposed a quantization and pruning framework for efficient RRAM IMC implementations.

\subsection*{Mapping Optimization}
Mapping neural networks to IMC architectures is a complex problem. \cite{li} proposed an approach to optimize the mapping of multimodal neural networks to IMC hardware to improve throughput. \cite{gopalakrishnan} developed a methodology to design convolutional neural networks that would map better to crossbar architectures. \cite{xiaochen} proposed a weight mapping methodology that would improve data reuse of convolutional layers on crossbars. \cite{kanghe} proposed a methodology to replicate layers in a neural network based on area freed-up by optimization of peripheral circuitry.

\subsection*{Optimization of peripheral circuitry}
The peripheral circuits of crossbar tiles i.e., the DAC and ADC systems are crucial parts of IMC designs. ADCs contribute to a large portion of the power and area budgets, and are thus, a major bottleneck in the design of IMC systems. Various optimized IMC designs have been proposed that address these bottlenecks. \cite{jiang} discusses an ADC design that implements shifts and adds in the analog domain. \cite{Saxena_2022} proposed replacing ADCs with 1-bit sense amplifiers and training neural networks to be tolerant to such aggressive partial sum quantization. 
\cite{kanghe} proposed an approach of decreasing the row parallelism to reduce the area overhead of ADCs and thus, improve the effective density of IMC chips.

To the best of our knowledge, LRMP is the first work that proposes a synergistic methodology that combine the benefits of mixed precision quantization and mapping optimization to jointly optimize the performance and accuracy of IMC-based neural network accelerators. LRMP is also the first work that proposes a linear programming-based approach to perform layer replication in IMC systems. Furthermore, circuit optimizations of tile peripheral are largely complementary to LRMP, and can be used to further improve the performance.
\section{Conclusion}
\label{sec:conclusion}

In-memory computing is a promising technology for accelerating neural networks by performing memory-intensive vector matrix multiplications using analog principles in a memory array. We propose LRMP, a method to synergistically perform layer replication and mixed precision quantization to improve performance of DNNs when mapped to area-constrained IMC accelerators. Our experiments suggest that LRMP can achieve considerable improvements in latency, throughput and energy consumption under iso-utilization and iso-accuracy constraints compared to 8-bit fixed point implementations. 

\bibliographystyle{IEEEtran}
\bibliography{references.bib}
\end{document}